\documentstyle[preprint,revtex,12pt]{aps}
\renewcommand{\baselinestretch}{1.0}
\begin{document}
\begin{title}
Nonequilibrium Dynamics and Aging in the\\
Three--Dimensional Ising Spin Glass Model
\end{title}
\renewcommand{\thefootnote}{\fnsymbol{footnote}}
\addtocounter{footnote}{1}\footnotetext{Present address: Institut f\"ur
Theoretische Physik, Universit\"at zu K\"oln, 5000 K\"oln 41, Germany.}
\author{Heiko Rieger\fnsymbol{footnote}}
%
%\author{Heiko Rieger}
%
\begin{instit}
Physics Department,
University of California,
Santa Cruz, CA 95064, USA\\
HLRZ c/o KFA J\"ulich,
Postfach 1913,
5170 J\"ulich, Germany
\end{instit}

\vskip1cm

\begin{abstract}
The low temperature dynamics of the three dimensional Ising spin glass
in zero field with a discrete bond distribution is investigated
via MC simulations. The thermoremanent magnetization
is found to decay algebraically and the temperature dependent exponents
agree very well with the experimentally determined values.
The nonequilibrium autocorrelation function $C(t,t_w)$ shows a crossover
at the waiting (or {\em aging}) time $t_w$ from algebraic {\em
quasi-equilibrium} decay for times $t$$\ll$$t_w$ to another,
faster algebraic decay for $t$$\gg$$t_w$ with an exponent similar
to one for the remanent magnetization.
\end{abstract}
%\pacs{75.10N, 75.50L, 75.40G.}
%
\narrowtext
\newpage
\baselineskip24pt
The measurement of dynamical nonequilibrium quantities in real spin glasses
\cite{BY} has a long history. The typical
experiment that has been conducted many times \cite{EXP1,EXP2} is the
following: Within a magnetic field the spin glass (for instance Cu(Mn),
Au(Fe), Fe$_{0.5}$Mn$_{0.5}$TiO$_3$,
(Fe$_x$Ni$_{(1-x)}$)$_{75}$P$_{16}$B$_6$Al$_3$,
Cd$_x$Mn$_{(1-x)}$Te, etc.)
is cooled down to temperatures below the freezing
temperature $T_g$ and either immediately or after a certain waiting time
$t_w$ the field is switched off. Then the so called (thermo)remanent
magnetization $M_{\rm rem}(t)$ is measured as a function of time $t$.
The asymptotic time dependence of this quantity is found to be algebraic
well below $T_g$ ($T/T_g\le0.98$) in the short range Ising spin glass
Fe$_{0.5}$Mn$_{0.5}$TiO$_3$ \cite{ito,nordito} and in an amorphous
metallic spin glass (Fe$_x$Ni$_{(1-x)}$)$_{75}$P$_{16}$B$_6$Al$_3$
\cite{REMEXP}. Furthermore the
time--dependence of the remanent magnetization depends on the waiting
time $t_w$, a phenomenon called aging \cite{EXP1}.

Several attempts have been made to explain this behavior theoretically
\cite{PSAA,KoHi,FiHu,Sibani,Bouch} and a wide variety of functional forms for
the time dependence of the remanent magnetization is found. The problem
lays in the fact that starting from a microscopic model or
model--Hamiltonian one encounters insurmountable difficulties in trying to
solve the nonequilibrium dynamics. Therefore additional assumptions have to
be made and the final outcome --- stretched exponential \cite{PSAA},
algebraic \cite{Sibani,Bouch} or logarithmic \cite{FiHu} decay ---
depends on them.
Even within the mean field approximation it is hard
to obtain any analytical \cite{GaDeMo,Horner} or semianalytical
\cite{Opper} results.

Once a microscopic model for a spin glass has been formulated, one can in
principle try to extract its macroscopic behavior via
Monte Carlo (MC) simulations. In contrast to an analytical treatment, where the
calculation of dynamical nonequilibrium quantities within the spin glass phase
(instead of those characterizing equilibrium, see \cite{Zipp}) is even more
complicated, MC simulation can be done with less effort for certain
nonequilibrium situations, since equilibration times reach astronomical
values in case of spin glasses \cite{BhaYou,OgMo,Og}. The remanent
magnetization with zero waiting time has been investigated numerically
for the mean field version of a spin glass model \cite{FiHu,Ki} and
for the three dimensional EA (Edwards--Anderson) model only right
at the critical temperature\cite{Huse}. Quite recently attempts have been
made to investigate the whole temperature range below $T_g$
numerically \cite{JOA,ritort}. However, a
systematic MC study of nonequilibrium correlations and aging phenomena
within the frozen phase ($T<T_c$) of the three--dimensional EA spin
glass model has not been made up to now. The results such an investigation,
its theoretical implications and comparison with experiments will be
reported in this letter.

The system under consideration is the three--dimensional Ising spin--glass
with nearest neighbor interactions and a discrete bond distribution. Its
Hamiltonian is
\begin{equation}
{\cal H} = -\sum_{\langle ij\rangle} J_{ij}\sigma_i\sigma_j\;,\label{hamilt}
\end{equation}
where the spins $\sigma_i=\pm1$ occupy the sites of a $L\times L\times L$
simple cubic lattice with periodic boundary conditions and the random
nearest neighbor interactions $J_{ij}$
take on the values $+1$ or $-1$ with probability $1/2$. We consider
single spin flip dynamics and used a special, very fast implementation
of the Metropolis algorithm on a Cray YM-P (see ref.\ \cite{mspin} for
details). The simulations were done in the frozen phase, that means at
temperatures below $T_c=1.175\pm0.025$ (see ref.\cite{BhaYou,OgMo,Og}).
All measured quantities are
averaged over at least 128 samples (smaller system sizes were averaged over
up to 1280 samples). The system size was increased until no further size
dependence of the results has been observed within the simulation time
($t\le10^6$), which is measured in MC sweeps through the whole
lattice. It turns out that $L=32$ is large enough for this time range
(cf.\ \cite{Og}).

The system was prepared in a fully magnetized initial configuration
and then the simulation was run
for a time $t_w$ (=waiting time) and then the spin configuration
$\underline{\sigma}(t_w)$ was stored. From now on after each MC
step (data were then averaged over appropriate time intervals, cf.\ \cite{Og})
the following correlation function was measured:
\begin{equation}
C(t,t_w)={1\over N}\sum_i\overline{
\langle\sigma_i(t+t_w)\sigma_i(t_w)\rangle}\;,\label{corr}
\end{equation}
where $\langle\cdots\rangle$ means a thermal average (i.e.\ an average
over different realizations of the thermal noise, but the same initial
configuration) and the bar means an average over different realizations of
the bond--disorder. The quantity $C(t,0)$ corresponds to the remaining
magnetization of the system after a time $t$.
\begin{equation}
M_{\rm rem}(t)=C(t,0)\;.\label{rem}
\end{equation}
This quantity is directly related to the experimentally determined
thermoremanent magnetization with zero waiting time (i.e.\ without aging)
and nearly saturated initial magnetization.
The result for the remanent magnetization $M_{\rm rem}(t)$ is shown in fig.\
\ref{fig1}
within a log--log plot. Its decay clearly obeys a power law for large times
and temperatures in the range $1.1\ge T\ge0.5$. The exponent $\lambda(T)$
for the fit
\begin{equation}
M_{\rm rem}(t)\propto t^{-\lambda(T)}\;,\quad(t\ge10^2)\label{remexp}
\end{equation}
is plotted in fig.\ \ref{fig2}, upper curve. It starts at
$\lambda=0.36\pm0.01$ for $T=1.1$ (and can be extrapolated via the
fit indicated in figure 2 to 0.39$\pm$0.01 for T=T$_c$, which was
already found in \cite{Huse}) and decreases monotonically with temperature.
For the short range Ising spin glass Fe$_{0.5}$Mn$_{0.5}$TiO$_3$ and for
certain amorphous metallic spin glasses not only the same algebraic decay
of the remanent magnetization has been observed, but also the shape of the
functional temperature dependence of the exponent $\lambda(T)$ and even
its numerical values are in excellent agreement: from figure 4b in ref.\
\cite{REMEXP} one may for instance read off $\lambda(T_g)\approx0.38$ and
$\lambda(0.5\,T_g)\approx0.12$, concurring within the errorbars
to the corresponding data plotted in fig.\ 2.

It was
argued \cite{FiHu} that the decay of $M_{\rm rem}(t)$ should be logarithmic
(i.e.\ $M_{\rm rem}(t)\propto(\ln t)^{-\lambda/\psi}$) below $T_c$, but a
fit of the data in fig.\ 1 for $T\ge0.5$ does not yield acceptable results
over the range of the observation time. We want to focus some attention to the
$T=0.4$ curve: It bends upward in the log--log plot for $t>10^4$, which
could indicate the onset of a slowlier than algebraic decay for
temperatures smaller than $0.5$, logarithmic for instance.

Another indication that something new happens at lower temperatures can be
obtained by looking at the short time behavior of $M_{\rm rem}(t)$: At
$T\approx0.5$ a plateau begins to develop for $t<10^2$, which can clearly
be seen for $T=0.4$ and becomes even more pronounced and wider for even
smaller temperatures. It can be excactly reproduced in shape and location
for smaller and larger sizes and number of samples, which means that
it is a physical effect and not only a fluctuation. A possible interpretation
might be that the system gets trapped in metastable states, whose lifetime
grows with decreasing temperatures.

Next we turn our attention to the correlation function $C(t,t_w)$ with
$t_w=10^a$ ($a=1,\ldots,5$). In contrast to $M_{{\rm rem}}(t)$ the
correlation function $C(t,t_w)$ for $t_w\ne0$ is not directly related
to the thermoremanent
magnetization $M(t,t_w)$ at time $t+t_w$ in a temperature--quench experiment,
where the field $H$ is switched off at time $t_w$ after the quench (see e.g.\
\cite{WAITEXP}). In equilibrium ($t_w\rightarrow\infty$) $C(t,\infty)$ is
related to the relaxation function $R(t,\infty)=M(t,\infty)/H$ via the
fluctuation dissipation theorem (FDT) $R(t,\infty)=C(t,\infty)/k_B T$.
However, in a nonequilibrium situation, like the one considered here,
slight differences between them exist \cite{JOA,egounpub} (e.g.\ in the
location of the maximum relaxation rate).
The magnetization that is induced by a small external field ($H\ll1$)
for model (\ref{hamilt}) is rather small, therefore the functional form of
$M(t,t_w)$ is harder to determine accurately via MC--simulations. This
is the reason why in this letter the focus is on $C(t,t_w)$.

A typical set of data for a particular
temperature ($T=0.8$) is shown in fig.\ \ref{fig3} in a log--log plot.
One observes a crossover from a slow algebraic decay for $t\ll t_w$ to a
faster algebraic decay for $t\gg t_w$. The crossover time is simply
defined as the intersection of the two straight line fits for short-- and
long--time behavior in the log--log plot. For the long--time behavior
the fit to
\begin{equation}
C(t,t_w)\propto t^{-\lambda(T,t_w)}\;,\quad t\gg t_w\label{correxp}
\end{equation}
yields a set of exponents that is depicted in fig.\ \ref{fig2}. For
increasing $t_w$ the exponent $\lambda(T,t_w)$ decreases only
slightly and the waiting time dependence becomes weaker for lower
temperatures. By looking at fig.\ \ref{fig3} one observes that it is
difficult to extract $\lambda(T,t_w)$ for $t_w=10^4$ and $10^5$ since there
are only 2 or 1 decades left to fit the exponent --- therefore they are not
shown in fig.\ 2. The exponent describing the short time ($t\ll t_w$)
behavior of $C(t,t_w)$,
\begin{equation}
C(t,t_w)\propto t^{-x(T)}\;,\quad t\ll t_w\;,\label{corrshort}
\end{equation}
which is depicted in fig.\ \ref{fig4}, is independent of the waiting time
$t_w$. Since the system was able to equilibrate over a time $t_w$, all
processes occuring on timescales smaller than $t_w$ have the charcteristics of
equilibrium dynamics and therefore the exponent $x(T)$ is identical to that
describing the decay of the equilibrium autocorrelation function
$q(t)=\lim_{t_w\rightarrow\infty} C(t,t_w)$. The latter was investigated in
\cite{Og} and the exponents that are reported there for $T\ge0.7T_c$ agree
with the values shown in fig.\ \ref{fig4}. They also agree with those
determined experimentally \cite{nordito} in the short range Ising spin glass
Fe$_{0.5}$Mn$_{0.5}$TiO$_3$ via the above mentioned relaxation function
$R(t,t_w)=M(t,t_w)/H$ for $t\ll t_w$ (note that in this
quasiequilibrium--regime $C(t,t_w)$ and $R(t,t_w)$ are related via the
FDT \cite{JOA,egounpub}, yielding the same exponents for both): close to
$T_g$ ($T/T_g=1.029$) they obtain $x=0.07$. Furthermore there seems
to be a temperature at about $0.3$, where $x(T)$ becomes zero, which could be
another indication for the above mentioned onset of a logarithmic decay of
the correlation functions \cite{FiHuEq}.

Although the decay of the
nonequilibrium correlations in the temperature range $0.5\le T\le1.1$ is
algebraic rather than logarithmic as predicted by the droplet picture
proposed in \cite{FiHu}, this picture might not be inappropriate:
Let us assume the following scaling law for the dependence of the free energy
barriers $B$ on a length scale $L$ of the regions to be relaxed: $B\propto
\Lambda\,{\rm ln} L$ instead of $B\propto L^\psi$ as in \cite{FiHu}. Then one
ends up with an algebraic decay of e.g.\ the remanent magnetization by
observing (see \cite{FiHu}) that the typical length  scale of domains $R_t$
now grows with time according to $\Lambda\,{\rm ln} R_t\sim T\,{\rm ln} t$,
which means $R_t\propto t^{T/\Lambda(T)}$, leading to equations
(\ref{rem}--\ref{corrshort}).

In the context of the phenomenological model for the dynamics and aging
in disordered systems developped in ref.\ \cite{Bouch}, the algebraic
decay of correlations found so far implies that the probability distribution
of free energy barriers is exponential in the temperature range of
$0.5\le T\le1.1$ for the system under consideration. Furthermore we would
like to mention that a fit to the functional form for the short time
behavior ($t\ll t_w$) $C(t,t_w)\sim 1-a(t/t_w)^y$ proposed in \cite{Bouch}
works also quite well for our data, although not as
convincingly as equation (\ref{corrshort}).

Guided by equations (\ref{remexp}) and (\ref{correxp})
we tried to put our results into the following scaling form:
\begin{equation}
C(t,t_w)=c_T\,t^{-x(T)}\,\Phi_T(t/t_w)\;,\label{scale}
\end{equation}
where $\Phi_T(y)=1$ for $y=0$ and
$\Phi_T(y)\propto y^{x(T)-\tilde{\lambda}(T)}$ for $y\rightarrow\infty$.
The form (\ref{scale}) has recently been used \cite{BhaYou2}
successfully to extract the critical dynamical exponent $z$ from the
nonequilibrium correlation function (\ref{corr}) via finite size scaling,
where the waiting time $t_w$ has been replaced by the relaxation
time $\tau\propto L^z$ in the critical region.
For temperatures below $T=0.8$ equation (\ref{scale}) yields an
acceptable fit (which can already be deduced from the neglegible waiting
time dependence of the exponents $\lambda(T,t_w)$ for $T\le0.7$, see fig.\
2).

Concluding we have reported new results of numerical nonequilibrium
simulations that show an excellent concurrence with experiments on the
short range Ising spin glass Fe$_{0.5}$Mn$_{0.5}$TiO$_3$ and
on amorphous metallic spin glasses: not only a single exponent but a
whole continuum of (temperature dependent) exponents for the remanent
magnetization are found to agree within the numerical errors.
Although the values for the exponents extracted from experiments
might vary somewhat depending on the microscopic details (range of
interactions, spin--type) the main features of the relaxation and
the dynamics of many different three--dimensional spin glasses are very
similar and the functional forms of the remanent magnetization decay
should be the same for different systems \cite{nordpriv}.

Furthermore we have shown that aging
phenomena in the spin glass model under consideration can be observed
via the measurement of a particular correlation function and that
its nonequilibrium dynamics is indeed gouverned by its equilibrium
characteristics for time scales smaller than the imposed waiting (or aging)
time. This gives an interesting new perspective (see also \cite{BhaYou2,Bray})
to extract equilibrium quantities, which are hard to obtain via MC simulations
within the spin glass phase. Finally, by observing
plateaus in the short time behavior and slowing down of the algebraic decay
of the remanent magnetization, we revealed a dynamical scenario
at very low temperatures that is not yet fully understood.

The author would like to thank A.\ P.\ Young for many extremely
valuable discussions. He is grateful to J.~O.~Andersson, D.~Belanger,
J.~P.~Bouchaud and P.~Nordblad for various comments, hints, suggestions
and explanations. The simulations were performed on the CRAY Y-MP
at the supercomputer center in J\"ulich and took about 100 hours of
CPU--time. Financial support from the DFG (Deutsche Forschungsgemeinschaft)
is also acknowledged.

\newpage

\figure{\label{fig1}
The remanent Magnetization $M_{\rm rem}(t)$
versus the time $t$ in a log--log plot
for varying temperatures. From top to bottom we have $T$=0.4, 0.5,
0.6, 0.7, 0.8, 0.9, 1.0 and 1.1. The sytem size is L=32
and the data are averaged over 128 samples. The errorbars are of the
size of the symbols for $M_{\rm rem}\le0.01$ and much smaller
for larger $M_{\rm rem}$.}

\figure{\label{fig2}
{\bf Upper curve:} The exponent $\lambda(T)$ for the remanent magnetization
(\ref{rem}) versus temperature. The points represented by diamonds
($\diamond$, plus errorbars) are those extracted from figure \ref{fig1}
and the full curve is a least square fit to a quadratic polynomial as a
guideline to the eye.\\
{\bf Lower Points:}
The nonequilibrium exponent $\lambda(T,t_w)$ (see equation
(\ref{correxp})) extracted from the long-time behavior ($t\gg t_w$)
of the nonequilibrium correlation function $C(t,t_w)$ (see figure
\ref{fig3}) for fixed values of $t_w$ versus the temperature $T$.
{}From top to bottom we have: ($\triangle$) $t_w$=10, ($\Box$) $t_w$=100
and ($\circ$) $t_w$=1000. The errorbars are indicated.}

\figure{\label{fig3}
The averaged nonequilibrium spin autocorrelation function
$C(t,t_w)$ of equation (\ref{corr})
for fixed values of $t_w$ versus time $t$ on a double logarithmic
time--scale. The temperature is fixed to be T=0.8 and from
top to bottom we have $t_w$=10$^5$, 10$^4$, 10$^3$, 10$^2$ and 10.
The system size is L=32 and the data are averaged over 128 samples.
The size of the errorbars is only a fraction of the size of the symbols.}

\figure{\label{fig4}
The equilibrium exponent $x(T)$ for the equilibrium autocorrelation
funcion $q(t)$ extracted from the short--time behavior ($t\ll t_w$) of
the nonequilibrium correlation function $C(t,t_w)$ (see equation
(\ref{corrshort})) versus the temperature $T$. The errorbars are smaller
than the circles, as indicated. For $T\le0.8$ the data are fitted to a
straight line, which shows that at approximately $T=0.3$ the exponent
$x(T)$ vanishes.}
\vfill
\eject
%
%***************************************************************************
%*                                                                         *
%*  Figure 1                                                               *
%*                                                                         *
%***************************************************************************
%
% GNUPLOT: LaTeX picture
\setlength{\unitlength}{0.240900pt}
\ifx\plotpoint\undefined\newsavebox{\plotpoint}\fi
\sbox{\plotpoint}{\rule[-0.175pt]{0.350pt}{0.350pt}}%

\end{center}
\vskip5cm
\begin{center}
\Large Fig.\ 4
\end{center}

\begin{references}

%\bibitem[*]{address} Present address: Institut f\"ur Theoretische Physik,
%Universit\"at zu K\"oln, 5000 K\"oln 41, Germany.
%
\bibitem{BY} For a review of spin glasses see
K.~Binder, and A.~P.~Young, Rev.\ Mod.\ Phys.\ {\bf 58},
801 (1986).

\bibitem{EXP1} L.~Lundgren, R.~Svedlindh, P.~Nordblad and O.~Beckman,
Phys.\ Rev.\ Lett.\ {\bf 51}, 911 (1983).

\bibitem{EXP2} R.~V.~Chamberlin, G.~Mozurkevich and R.~Orbach,
Phys.\ Rev.\ Lett.\ {\bf 52}, 867 (1984);
R.~Hoogerbeets, Wei--Li Luo and R.~Orbach
Phys.\ Rev.\ B {\bf 34} 1719, (1986);
M.~Alba, J.~Hamman, M.~Ocio, Ph.~Refregier and  H.~Bouchiat,
J.\ Appl.\ Phys.\ {\bf 61}, 3683 (1987);
M.~Ledermann, R.~Orbach, J.~M.~Hammann, M.~Ocio and E.~Vincent,
Phys.\ Rev.\ B {\bf 44}, 7403 (1991).


\bibitem{ito} A.~Ito, H.~Aruga, E.~Torikai, M.~Kikuchi, Y.~Syono
and H.~Takai, Phys.\ Rev.\ Lett.\ {\bf 57}, 483 (1986).

\bibitem{nordito} K.~Gunnarson, P.~Svedlindh, P.~Nordblad, L.~Lundgren,
H.~Aruga and A.~Ito, Phys.\ Rev.\ Lett.\ {\bf 61}, 754 (1988).

\bibitem{REMEXP} P.~Granberg, P.~Svedlindh, P.~Nordblad, L.~Lundgren
and H.~S.~Chen, Phys.\ Rev.\ B {\bf 35}, 2075 (1987).

\bibitem{PSAA} R.~G.~Palmer, D.~L.~Stein, E.~Abrahams and P.~W.~Anderson,
Phys.\ Rev.\ Lett.\ {\bf 53}, 958 (1984).

\bibitem{KoHi} G.~J.~M. Koper and H.~J.~Hilhorst,
J.\ Phys.\ France {\bf 49}, 429 (1988).

\bibitem{FiHu} D.~S.~Fisher and D.~A.~Huse,
Phys.\ Rev.\ B {\bf 38}, 373 (1988).

\bibitem{Sibani} P.~Sibani and K.~H.~Hoffmann, Phys.\ Rev.\ Lett.\ {\bf 63},
2853 (1989).

\bibitem{Bouch} J.~P.~Bouchaud,
J.\ Physique I {\bf 2}, 1705 (1992);
J.~P.~Bouchaud, E.~Vincent and J.~Hammann, Preprint (1993).

%\bibitem{REMARK} Once the critical exponents of the equilibrium phase
%transition of the 3d EA model are known, some progress can be made
%along the lines of: H.~K.~Janssen, B.~Schaub and B.~Schmittmann,
%Z.\ Phys.\ B {\bf 73}, 529 (1989).

\bibitem{GaDeMo} E.~Gardner, B.~Derrida and P.~Mottishaw,
J.\ Phys.\ France {\bf 48}, 741 (1987);
M.~Schreckenberg and
H.~Rieger, Z.\ Phys.\ B {\bf 86}, 443 (1992).

\bibitem{Horner} H.~Horner, Z.\ Phys.\ B {\bf 78}, 27 (1990).

\bibitem{Opper} H.~Eissfeller and M.~Opper,
Phys.\ Rev.\ Lett.\ {\bf 68}, 2094 (1992).

\bibitem{Zipp} H.~Sompolinsky and A.~Zippelius,
Phys.\ Rev.\ Lett.\ {\bf 47}, 359 (1981).

\bibitem{BhaYou} R.~N.~ Bhatt and A.~P.~Young,
Phys.\ Rev.\ Lett.\ {\bf 54}, 924 (1985); Phys.\ Rev.\ B {\bf 37}, 5606
(1988).

\bibitem{OgMo} A.~T.~Ogielski and I.\ Morgenstern,
Phys.\ Rev.\ Lett.\ {\bf 54}, 928 (1985).

\bibitem{Og} A.~T.~Ogielski, Phys.\ Rev.\ B {\bf 32}, 7384 (1985).

\bibitem{Ki} W.~Kinzel, Phys.\ Rev.\ B {\bf 33}, 5086 (1986).

\bibitem{Huse} D.~Huse, Phys.\ Rev.\ B {\bf 40}, 304 (1989).

\bibitem{JOA} J.~O.~Andersson, J.~Mattson and P.~Svedlindh, Phys.\ Rev.\
B {\bf 46}, 8297 (1992).

\bibitem{ritort} G.~Parisi and F.~Ritort, Preprint ROM2F/92/40.

\bibitem{mspin} H.~O.~Heuer, Comp.\ Phys.\ Comm.\ {\bf 59}, 387 (1990);
H.~Rieger, J.\ Stat.\ Phys.\ {\bf 70}, 1063 (1993).

\bibitem{WAITEXP} P.~Nordblad, P.~Svedlindh, L.~Lundgren and
L.~Sandlund, Phys.\ Rev.\ B {\bf 33}, 645 (1986);\\
M.~Alba, M.~Ocio and J.~Hammann, Europhys.\ Lett.\ {\bf 2}, 45 (1986).

\bibitem{egounpub} H.~Rieger, unpublished.

\bibitem{FiHuEq} According to D.~S.~Fisher and D.~A.~Huse,
Phys.\ Rev.\ Lett.\ {\bf 56}, 1601 (1986), also the equilibrium
autocorrelation function $q(t)$ should decay logarithmically
$q(t)\propto({\rm ln}t)^{-\theta/\psi}$.

\bibitem{BhaYou2} R.~N.~ Bhatt and A.~P.~Young,
Europhys.\  Let.\ {\bf 20}, 59 (1992).

\bibitem{nordpriv} P.~Nordblad, private communication.

\bibitem{Bray} R.~E.~Blundell, K.~Humayun and A.~J.~Bray, J.\ Phys.\ A
{\bf 25}, L733 (1992).

\end{references}
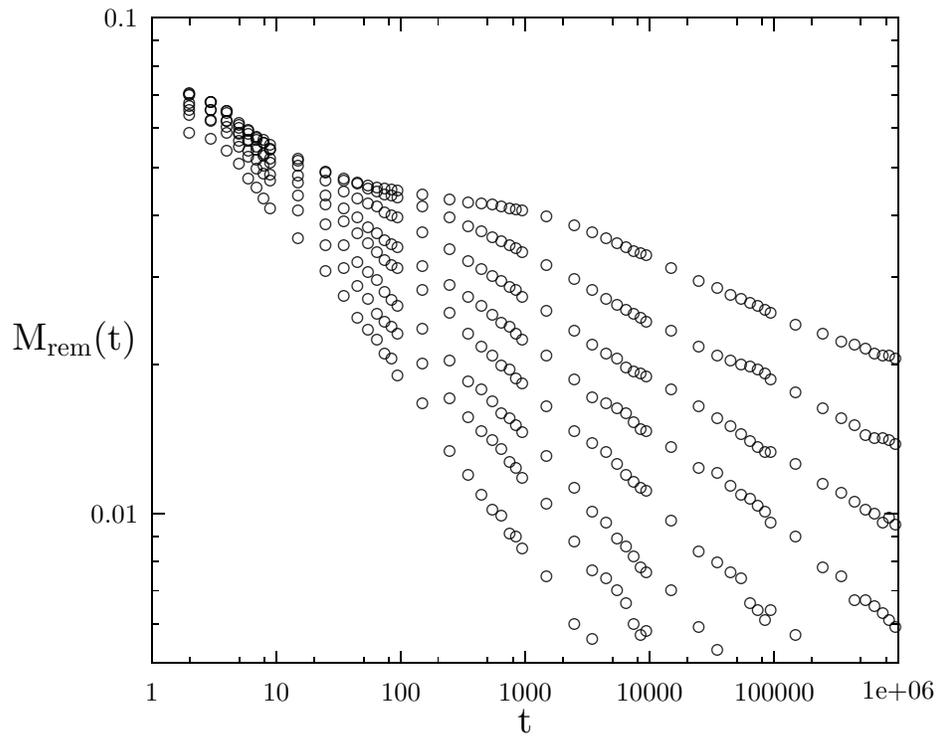
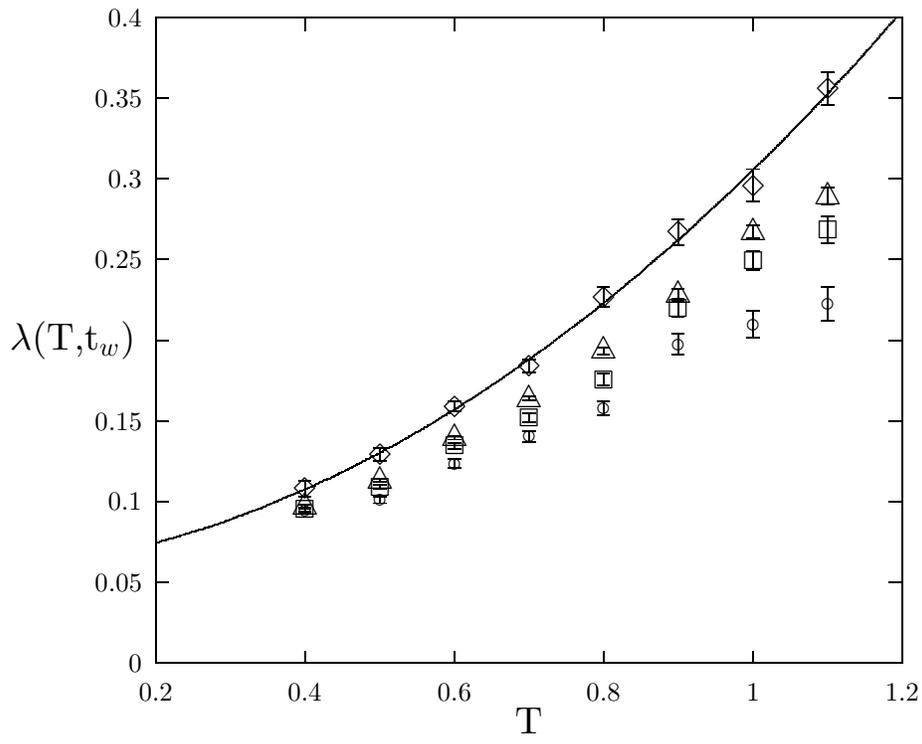
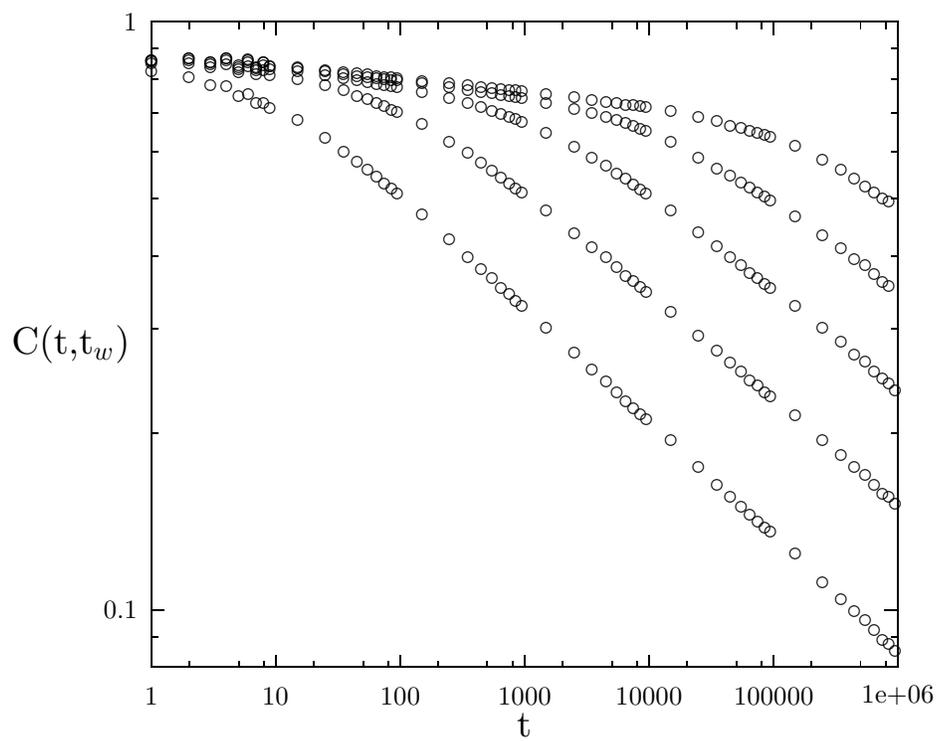
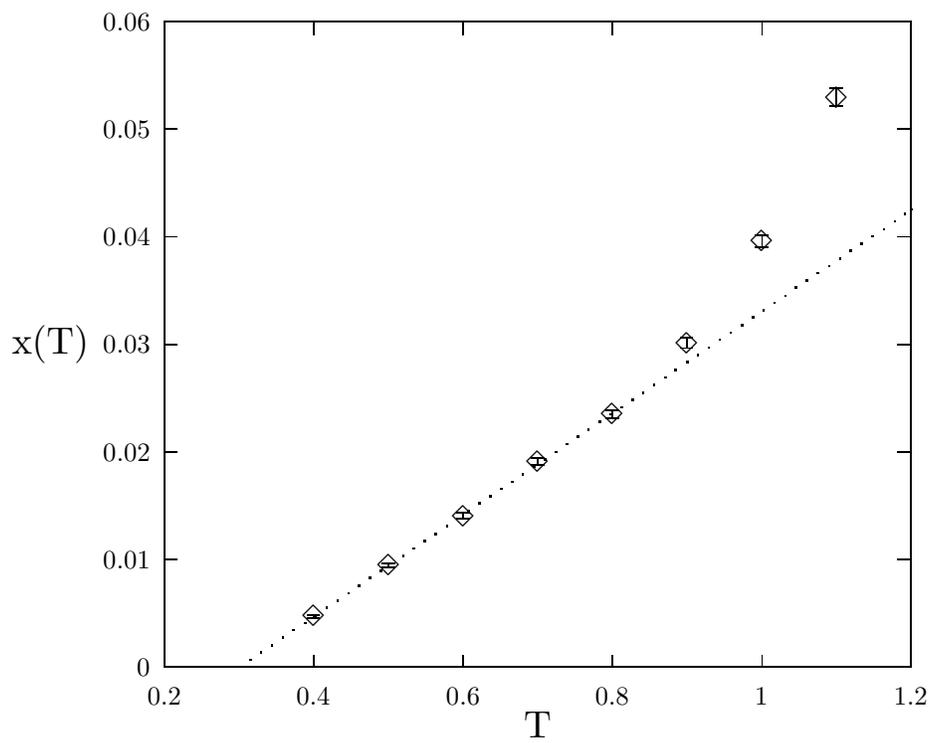
\end{document}